\documentclass[aps,pre,a4paper,twocolumn,nofootinbib,superscriptaddress]{revtex4-2}

\usepackage{graphicx}
\usepackage{hyperref}
\usepackage{latexsym}
\usepackage{amsmath, amssymb, amsthm}
\usepackage{mathtools}
\usepackage{xcolor}
\usepackage{comment}

\usepackage{microtype}
\usepackage{epigraph}
\usepackage{lipsum}

\usepackage[normalem]{ulem}

\newcommand{\change}[1]{{\color{black}{#1}}}

\newcommand{\km}{k_{\text{min}}}
\newcommand{\kcdop}{(k_a)_c^{\text{DOP}}}
\newcommand{\kccmp}{(k_a)_c^{\text{CMP}}}
\newcommand{\kac}{(k_a)_c^{\text{CMP}}}
\newcommand{\be}{\begin{equation}}
\newcommand{\ee}{\end{equation}}

\begin{document}


\title{Cumulative Merging Percolation: A long-range percolation process in networks}


\author{Lorenzo Cirigliano}

\affiliation{Dipartimento di Fisica Universit\`a ``Sapienza”, P.le
  A. Moro, 2, I-00185 Rome, Italy.}

\affiliation{Centro Ricerche Enrico Fermi, Piazza del Viminale, 1,
  I-00184 Rome, Italy}

\author{Giulio Cimini}
\affiliation{Dipartimento di Fisica e INFN, Università di Roma “Tor
  Vergata”, I-00133 Rome, Italy}

\affiliation{Centro Ricerche Enrico Fermi, Piazza del Viminale, 1,
  I-00184 Rome, Italy}

\author{Romualdo Pastor-Satorras}

\affiliation{Departament de F\'{\i}sica, Universitat Polit\`ecnica de
  Catalunya, Campus Nord B4, 08034 Barcelona, Spain}

\author{Claudio Castellano}

\affiliation{Istituto dei Sistemi Complessi (ISC-CNR), Via dei Taurini
  19, I-00185 Rome, Italy}

\affiliation{Centro Ricerche Enrico Fermi, Piazza del Viminale, 1,
  I-00184 Rome, Italy}


\date{\today}

\begin{abstract}
  Percolation on networks is a common framework to model a wide range
  of processes, from cascading failures to epidemic spreading.  Standard
  percolation assumes short-range interactions, implying that nodes can
  merge into clusters only if they are nearest-neighbors.  Cumulative
  Merging Percolation (CMP) is an new percolation process that assumes
  long-range interactions, such that nodes can merge into clusters even
  if they are topologically distant.
  Hence in CMP percolation clusters do not coincide with the topological
  connected components of the network.  Previous work has shown that a
  specific formulation of CMP features peculiar mechanisms for the
  formation of the giant cluster, and allows to model different network
  dynamics such as recurrent epidemic processes.  Here we develop a more
  general formulation of CMP in terms of the functional form of the
  cluster interaction range, showing an even richer phase transition
  scenario with competition of different mechanisms resulting in
  crossover phenomena. Our analytic predictions are confirmed by
  numerical simulations.
\end{abstract}


\maketitle

\section{Introduction}
Percolation theory is among the most developed fields of statistical
mechanics and mathematical physics.  A percolation process can be
defined as follows: We have a collection of elements and some
connections among them. This object is called a
\textit{graph}, the elements are called \textit{vertices} (or
\textit{nodes}) and the connections are called \textit{edges} (or
\textit{links}).  We then remove some nodes according to a certain
probabilistic or deterministic rule.  For instance, we can remove all
nodes with more than a given number of connections, or we can remove
nodes uniformly at random.  At the end of the removal process we wonder
about which connectivity properties are preserved.  The main problem of
percolation theory is to understand if a giant component (GC), that is a
connected component of extensive size, still exists in the graph after
the removal process.

Originally, percolation was studied on various types of low-dimensional
lattices.  The powerful methods of statistical mechanics of phase
transitions and critical phenomena, such as mean-field approximations,
renormalization group, asymptotic expansions and scaling theory, provide
us with a complete understanding of the percolation process on regular
lattice topologies~\cite{stauffer2018introduction}. In the past 20
years, the interest about complex networks has led to a great deal of
activity concerning percolation processes on
graphs~\cite{callaway2000network, Cohen2000, Dorogovtsev2008,
  Karrer2014,Li2021}.

Percolation processes are used to model a wide range of natural
phenomena, just by changing the underlying graph or the probabilistic
rule that determines the removal of nodes. For instance, percolation on
a regular lattice can model transport processes in porous media, such as
electrical and hydraulic conduction, air permeability and
diffusion~\cite{Sahini1994}.  On the other hand, percolation on random
graphs~\cite{newman2001random} can be used to investigate the robustness
of a networked system under intentional or random
attacks~\cite{callaway2000network}.

Additionally, a deep connection exists between percolation and epidemic
spreading. Indeed, the fundamental susceptible-infected-recovered (SIR)
epidemic model on networks~\cite{Kiss2017} can be mapped onto a bond
percolation process~\cite{Newman2002}. Such a mapping allows to use
the tools of percolation theory to get a full understanding of
static properties of the SIR model.  For epidemic processes which admit
a stationary steady state, such as the susceptible-infected-susceptible
(SIS) model, this mapping is less immediate, and has been realized only
through a new percolation model recently
proposed~\cite{menard2016percolation}.

Such a model, called cumulative merging percolation (CMP), is a truly
long-range percolation process.
This specification (long-range percolation) is often used for models
where, in a lattice, additional links connecting sites
separated by any euclidean distance are added, with a probability depending
on the distance~\cite{Zhang_1983,https://doi.org/10.1002/rsa.1022}.
In CMP instead, distances are only topological and two nodes can belong
to the same cluster even if no path of nearest-neighbor nodes connecting
them belongs to the cluster itself.
Models of similar type, called extended-neighborhood percolation models,
have been studied on regular lattices~\cite{Malarz2005,Malarz2015,Xun2020,Xun2021}.
In such models percolation clusters do not need to coincide with
topologically connected components, as two nodes may form a cluster even
if they are not nearest-neighbors but separated
by paths of {\em finite} length (typically 2 or 3).
In CMP this length can be arbitrarily large.

CMP has been recently studied in a specific {\em degree-ordered}
case~\cite{castellano2020cumulative} to elucidate the behavior of the
SIS model on random uncorrelated networks with power-law degree
distribution $P(k) \sim k^{-\gamma}$.  By means of a scaling approach
and numerical simulations, it was shown that the long-range nature of
the model guarantees, for any $\gamma>2$, the existence of a percolating
cluster for any value of the control parameter (i.e., the degree
threshold that determines node removal), at variance with what happens
for the short-range counterpart~\cite{Lee2013,caligiuri2020degree}.  The
aim of the present work is to define in full generality CMP on networks.
We present a scaling theory that extends the one introduced in
Ref.~\cite{castellano2020cumulative} to more general forms of the {\em
  interaction range}. We then consider in detail two paradigmatic
functional forms of the interaction range, deriving predictions
concerning the existence of a phase-transition at finite or infinite
value of the control parameter and the associated critical behavior.  We
obtain a rich scenario with competition of different mechanisms
resulting in crossover phenomena, which is confirmed by means of
numerical simulations.

The paper is organized as follows: In Sec.~\ref{sec:II}, we first define
the CMP process in the most general form and present the detailed
specifications considered in the rest of the paper.  In
Sec.~\ref{sec:III} we write a general scaling approach used to analyze
the process. Sec.~\ref{sec:IV} is dedicated to a detailed analysis of
two classes of CMP, characterized by algebraically and logarithmically
growing interaction range.  Finally, in Sec~\ref{sec:VI}, we summarize
our main results and present some possible future research paths.

\section{\label{sec:II}Cumulative Merging Percolation}

In classical (short-range) percolation, a cluster coincides with a
topologically connected component $C$, i.e., a subset of nodes such that
for any two nodes $i$ and $j$ there exists a path connecting them, made
of nearest-neighboring nodes belonging to $C$.  Thus, for instance, if
all nearest neighbors of a node are removed, this node cannot be a part
of any cluster -- except the one formed by itself only. In order to
introduce a long-range model, we need to go beyond such a definition. In
this section, we define a general procedure to define clusters that may
be composed of \textit{topologically disconnected and arbitrarily
  distant} components.

\subsection{General definition}

We denote a graph by $\mathcal{G}(V,E)$, where $V$ is the set of nodes
and $E$ is the set of edges.  Each node $i \in V$ is endowed with a
non-negative mass $m_i\geq 0$.
The mass of the set composed by two nodes $i$ and $j$ is given by
the sum of their masses
\be
m = m_i + m_j.
\label{mdef}
\ee

A \textit{partition} of the graph $\mathcal{P}(\mathcal{G})$ is a
collection of subsets of nodes
\begin{equation}
 \mathcal{P}(\mathcal{G})=\left\{ A\right\}_{A\subseteq V} \nonumber
\end{equation}
such that: (i) the sets in $\mathcal{P}$ cover $V$; (ii) every element
in $V$ belongs to exactly one subset in $\mathcal{P}$.  Each element of
$\mathcal{P}$ is called \textit{cluster}, denoting with $C_i$ the
cluster to which node $i$ belongs.  Notice that by definition each node
belongs only to one cluster.
Because of Eq.~\eqref{mdef}, the mass of a cluster is the sum of the
masses of all nodes belonging to it.
We define the \textit{interaction range}
of a cluster to be a non-decreasing function of its mass, $r(m)$.  We
stress that these clusters need not be topologically connected
components.

Given a pair of nodes $(i,j)$, the \textit{merging operator}
$M_{i,j}$, acting on the space of all partitions of the graph,
merges the two clusters $C_i$ and $C_j$ if and only if
\begin{equation}
  d_{i,j} \leq \min\{r(m_{C_i}),r(m_{C_j}) \};
\label{merging}
\end{equation}
where $d_{i,j}$ is the topological distance between $i$ and $j$.
If the condition~eqref{marging} is not fulfilled,
the merging operator leaves the two clusters unaltered. Notice that the
merging occurs only if node $i$ is within the interaction range of the
cluster $C_j$ and vice-versa, and that two clusters may be merged
together even if they are at arbitrary topological distance with each other,
provided the interaction ranges are sufficiently large.

We define the \textit{cumulative merging procedure} as follows:
\begin{enumerate}
  \item fix an infinite sequence $(i_t, j_t)_{t\in \mathbb{N}}$ of pairs of nodes of $V$;
  \item start from the finest partition $\mathcal{P}^0 \coloneqq
    \left\{ \{ i\}, i\in V \right\}$;
  \item iteratively apply the merging operator
  \begin{equation}
    \mathcal{P}^{t+1} = M_{i_t,j_t}(\mathcal{P}^t).
  \end{equation}
\end{enumerate}
The asymptotic partition
\begin{equation}
  \mathcal{P}^{\infty}\coloneqq \lim_{t\to \infty} \mathcal{P}^t,
\end{equation}
depends, in principle, on the sequence $(i_t, j_t)_{t\in \mathbb{N}}$.
Consider for instance the case in which all nodes have an infinite
interaction range, except node $a$ and node $b$, which are distant nodes
with small interaction range.  The sequence $(i_t,j_t)=(a,b)$ for every
$t$ leads to $\mathcal{P}^{\infty}=\mathcal{P}^0$, while it is immediate
to realize that $\mathcal{P}^{\infty}$ will be different for other
sequences.  A way to overcome this difficulty is to consider only
\textit{recurrent} sequences: we say that the sequence
$(i_t, j_t)_{t\in \mathbb{N}}$ is \textit{recurrent} if
$\left\{i_t,j_t \right\} = \left\{k,l\right\}$ infinitely many times,
for every $k,l \in V$ with $k\neq l$. An important result on the
asymptotic partition, whose proof can be found in
Ref.~\cite{menard2016percolation}, guarantees that the cumulative
merging procedure is well defined, i.e. sequence-independent, provided
that the sequence is recurrent.

In summary, given a graph, a collection of node masses, and the
function $r(m)$, the cumulative merging procedure
generates a unique partition of the graph in clusters, not necessarily
topologically connected.  Notice that nodes with $r(m_i) < 1$ by
construction do not play any role (i.e., they necessarily form clusters
of size 1), apart from determining the
topological distances among other nodes. We denote them as {\em inactive},
while nodes with $r(m_i) \ge 1$ are {\em active}, as they may participate
in merging events and form clusters of size larger than 1.

The CMP model defined above can be seen as a percolation process:
Depending on the function $r(m)$, the node masses and the
underlying graph, the asymptotic partition may be composed of
microscopic clusters only or include a giant cluster encompassing a
finite fraction of the total number of nodes.  It is a long-range
percolation model because clusters may be composed by different and
arbitrarily distant topologically connected components. This is
qualitatively different from extended-range percolation models which
allow only finite distances between disconnected components.

\subsection{A specific class of CMP models}

Following Refs.~\cite{menard2016percolation,castellano2020cumulative},
we set node masses to be equal to their degree, i.e. $m_i=k_i$.
In this way, the mass of a cluster is the sum of the degrees of the
nodes that belong to it.
Moreover, we take the interaction range to be a function
of the ratio between $m$ and a parameter $k_a$, that plays
the role of control parameter in the percolation problem,
\begin{eqnarray}
  r(m) = f(m/k_a),
  \label{rm}
\end{eqnarray}
where $f(x)$ is a non-decreasing function, with $f(1)=1$ and $f(x)<1$
for $x<1$.  According to this definition all nodes with degree $k_i<k_a$
have interaction range $r<1$ and are thus inactive.
Nodes with degree larger than or equal to $k_a$ are active.

\begin{figure}[t]
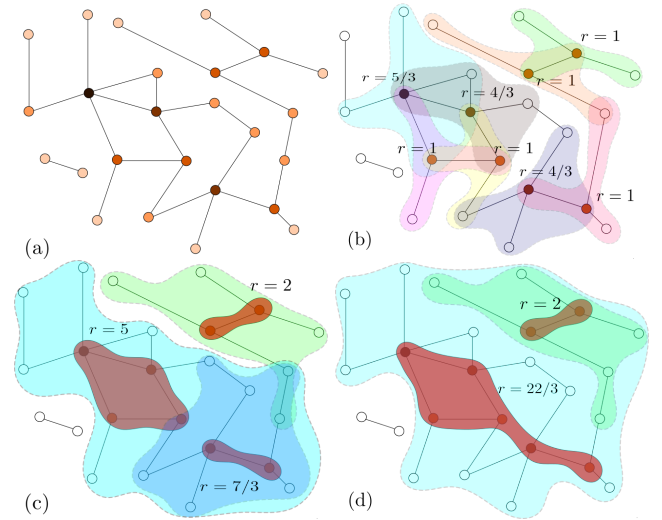

  \centering
  \includegraphics[width=0.23\textwidth]{CMP_0}
  \includegraphics[width=0.23\textwidth]{CMP_2}\\
  \includegraphics[width=0.23\textwidth]{CMP_3}
  \includegraphics[width=0.23\textwidth]{CMP_4}
  \caption{Visual representation of a CMP process on a graph with
    $k_a=3$ and $r(m)=m/k_a$.  (a) All nodes in the graph are
    shown. Colors depend on the degree $k$.  (b) Initial configuration
    of the merging process, with $k_a=3$, each node forming a cluster.
    Empty circles are inactive nodes.  Colored dashed regions represent
    the interaction range of each active node.  (c) Intermediate
    configuration of the merging process. Dark red regions represent
    clusters. (d) Final configuration of the merging process. From (c)
    to (d) the long-range nature of the process plays a crucial
    role. Note that the process ends since the interaction range of the
    cluster with $r=2$ does not reach any node of the cluster with
    $r=22/3$.}
  \label{CMP_example}
\end{figure}

In this way percolation occurs in a \textit{degree-ordered} way:
Increasing the control parameter $k_a$ is equivalent to removing nodes of
increasingly higher degree.  For the minimal value $k_a=\km$, all nodes
are active and hence the clusters coincide with the components of the
underlying graph.  For large $k_a \to \infty$ the number of active nodes
gets smaller and smaller and the nontrivial question is whether an
extensive cluster still exists.
With all these specifications, given a network and the value of $k_a$,
the final partition of the CMP is univocally defined.
Fig.~\ref{CMP_example} reports an example of how the CMP process
unfolds and what is the final partition of the graph.

Other choices for the initial masses, the interaction range and the
underlying graph are possible,
leading to a wealth of different models with different
critical properties.  For example, if $r(m)=1$, for any $m \geq 1$, CMP
coincides with standard site percolation, either in its degree-ordered
version~\cite{Lee2013,caligiuri2020degree}, if the initial mass of node
$i$ is $m_i=k_i \Theta(k_i-k_a)$, or in its random version, if
$m_i=k_i$ with probability $\phi$ and 0 otherwise.
Taking instead a generic interaction range $r(m)$ and again
$m_i=k_i$ with probability $\phi$ and 0 otherwise,
one has a truly long-range CMP with random activation.
The investigation of these and other variants constitutes an interesting
avenue for future research.

In the following we study this class of degree-ordered CMP processes
on power-law degree-distributed networks, described by uncorrelated
random graphs with degree distribution, in the continuous approximation,
\begin{equation}
  P(k)=(\gamma-1)\km^{\gamma-1}k^{-\gamma}
\end{equation}
where $\gamma>2$ and $\km$ is the minimum degree of nodes in the network.

Our aim is to investigate the formation of a CMP giant cluster
(CMPGC)~\footnote{This object was called CMP giant component in
  Ref.~\cite{castellano2020cumulative}.  We prefer to change
  denomination here to stress the difference between topologically
  connected components and CMP clusters.}.  Hence we focus on the
quantity $S_{\text{CMP}}$, defined as the fraction of nodes belonging to
the CMP largest cluster, as a function of the control parameter $k_a$.
In percolation theory it is customary to consider the fraction of active
(non removed) nodes, namely $\phi$, as control parameter.  In the large
$N$ limit,
\begin{equation}
\phi = \frac{N_a}{N} = \int_{k_a}^{\infty}dk P(k) =
\left(\frac{k_a}{\km} \right)^{1-\gamma},
\end{equation}
where $N_a$ is the number of active nodes.  From this equation, it
follows that we can express $S_{\text{CMP}}$ as a function of $\phi$ by
just replacing $k_a/\km$ with $\phi^{1/(1-\gamma)}$. In particular, the
behavior for large $k_a$ corresponds to the behavior for small fraction
$\phi$ of active nodes. Defining $\phi_c$ as the threshold value at
which a macroscopic CMPGC first appears, we expect that, close to the
transition
\begin{eqnarray}
  \label{eq:2}
  S_{\text{CMP}} \sim (\phi-\phi_c)^{\beta},
\end{eqnarray}
where $\beta$ is a characteristic exponent.  In the case $\phi_c=0$, we
expect
\begin{equation}
  \label{eq:3}
  S_{\text{CMP}} \sim \phi^{\beta} \sim k_a^{\beta(1-\gamma)},
\end{equation}
that is, a decay of the CMPGC size as a function of
$k_a$.

\section{General scaling theory for  degree-ordered CMP}

\label{sec:III}

\subsection{CMP for $\gamma\leq 3$}

A simple observation allows us to characterize the behavior of
$S_{\text{CMP}}$ on networks with $2<\gamma\leq 3$. The short-range
counterpart, i.e. with $r(m)=1$, of the CMP as defined above is called
\textit{degree-ordered percolation} (DOP). Its behavior on power-law
distributed networks has been studied in
Refs.~\cite{Lee2013,caligiuri2020degree}, where a detailed investigation
of the critical properties of the DOP giant component (DOPGC) can be
found.  In particular, it has been shown that:
\begin{itemize}
  \item for $\gamma\leq 3$ a DOPGC always exists for any finite value of $k_a$;
  \item for $\gamma>3$, a DOPGC exists only up to a finite critical point $\kcdop$.
\end{itemize}

From this result we can infer that for $\gamma \leq 3$ a CMPGC always
exists for every value of $k_a$. Indeed, since the DOPGC is always a
subset of the CMPGC, it follows that
$S_{\text{DOP}} \leq S_{\text{CMP}}$ for every $k_a$, and since the
short-range model has an infinite critical point, so has the long-range
model.  Furthermore, essentially all active nodes belong to the
DOPGC~\cite{castellano2020cumulative} and thus
\begin{equation}
 S_{\text{CMP}} \simeq \left(\frac{k_a}{\km}\right)^{1-\gamma}.
\end{equation}
Notice that this result is valid in full generality as long as $r(m)$ is
a non-decreasing function.  Instead, for $\gamma>3$ the DOP has a
transition to a phase with no giant component at $\kcdop$ and this does
not allow us to draw any conclusions a priori on $S_{\text{CMP}}$ for
large $k_a$.  In the following section, we develop a general scaling
theory to understand the behavior of $S_{\text{CMP}}$ for $\gamma>3$.

\subsection{CMP for $\gamma>3$}

Following~\cite{castellano2020cumulative}, we identify two different
mechanisms that may contribute to the formation of a CMPGC:
\begin{enumerate}
  \renewcommand{\labelenumi}{A:}
\item Extended DOP mechanism:
  it is essentially an extension of the DOP process involving the merging
  of DOP clusters separated by distances larger than $1$;
    \renewcommand{\labelenumi}{B:}
  \item Merging of distant isolated nodes: it works for high values of
    $k_a$ when essentially all active nodes are isolated and, on average,
    at large distance from each other.
\end{enumerate}

We compute the scaling of the order parameter $S_{\text{CMP}}$ with
$k_a$ due to each mechanism separately, and we identify the ranges of
$k_a$ values where one of them dominates over the other.

\subsubsection{Extended DOP mechanism}

In the DOP model for $\gamma>3$ there is an extensive giant cluster for
$k_a$ up to $\kcdop$. Above this threshold no DOP cluster is
extensive. In the CMP model, due to the interaction range extending
beyond 1, DOP clusters may merge and this may lead to the formation of a
CMPGC even for $k_a>\kcdop$. The most natural candidates for this
merging are small clusters at distance $2$ from each other or massive
isolated nodes at distance $2$ from small clusters.  We consider these
two contributions separately.

If two small clusters are topologically isolated
but at distance 2 from each other, the two clusters merge
provided their interaction range is larger than their relative
distance.
The merging of small clusters at distance 2 may then make possible
the merging of other, more distant small clusters and so on.
Assuming that in this way all $N_{\text{NI}}$ topologically nonisolated
nodes enter the CMPGC, this mechanism gives a largest cluster of size
(see Ref.~\cite{castellano2020cumulative} for details)
\begin{align}
  \frac{N_{\text{NI}}}{N} &\simeq \left(\frac{k_a}{\km} \right)^{1-\gamma}
  \left[\frac{\gamma-1}{\gamma-2}\km \left(\frac{k_a}{\km}\right)^{3-\gamma} \right] \nonumber \\
  &= \langle k \rangle \left(\frac{k_a}{\km}\right)^{2(2-\gamma)}.
  \label{eq:nonisolated}
\end{align}

The plausibility of this assumption strongly depends on the shape of the
function $f(x)$ in Eq.~\eqref{rm} and on the value of $k_a$.  If $f(x)$
grows as $x$ or faster, a cluster of two nodes has by definition an
interaction range equal at least to 2.  Hence such a cluster can merge
if at distance 2 from another cluster.  If instead $f(x)$ grows more
slowly the interaction range of a cluster formed by two adjacent nodes
may be smaller than 2, and the extended DOP mechanism becomes
ineffective.  In such a case Eq.~\eqref{eq:nonisolated} is a rough
overestimation of the largest cluster size.  In addition, for small
$k_a$ there are many active nodes and it is plausible that distances
among connected clusters are small.  For large $k_a$ instead, clusters
will tend to be at larger distances from each other, and an interaction
range equal to 2 will not be sufficient to guarantee the merging.

In a similar way, we can argue that also an isolated node can merge with
a DOP cluster at distance 2 from it, if it is massive enough so that its
interaction range is at least 2.  If we define $k_e$ the value of $k$ such
that $r(k_e) = 2$, and assume that all isolated nodes with degree
$k \geq k_e$ become in this way part of the CMPGC, we can estimate
$S_{\text{CMP}}$ as the fraction of isolated nodes with
$k \geq k_e$.
Following~\cite{castellano2020cumulative}, considering that $k_e$ is
proportional to $k_a$, and taking the limit $k_a \gg \km$, we
have~\cite{castellano2020cumulative}
\begin{equation}
  \frac{N_{r\geq 2}}{N} \sim \left(\frac{k_e}{\km}
  \right)^{1-\gamma}\left[1-\frac{\gamma-1}{\gamma-2}\frac{k_e}{k_c}
  \right],
\end{equation}
an expression that is also a clear overestimate of the true
contribution, getting worse for large $k_a$.

Summing up the two contributions we obtain
\begin{equation}
  S_{\text{CMP}}^{(1)} \simeq \frac{N_{\text{NI}}}{N}+\frac{N_{r\geq2}}{N}.
  \label{eq:size_1}
\end{equation}
Since the decay of $N_{r\ge 2}/N$ with $k_a$ is slower than the decay of
$N_{\mathrm{NI}}/N$, the first term in Eq.~\eqref{eq:size_1} dominates,
in principle, only up to a crossover scale $k_1^*$, which depends on
$\gamma$ and on the detailed functional form of the interaction range.
However, as noticed in~\cite{castellano2020cumulative}, one does not
expect to actually observe such a crossover as for asymptotically large
$k_a$ this mechanism cannot be active.  Indeed, for small values of
$k_a$ (many active nodes), distances between small clusters are
typically small, favoring the merging process.  As $k_a$ is increased,
in particular beyond $\kcdop$, typical distances between active nodes
grow larger and the extended DOP mechanism is strongly suppressed.
Therefore, we can assume the contribution of the extended DOP mechanism
to the size of the CMPGC to be
\begin{equation}
S_{\text{CMP}}^{(1)} \simeq \langle k \rangle \left(\frac{k_a}{\km}\right)^{2(2-\gamma)}
\label{eq:SCMP1}
\end{equation}
up to a finite value of $k_a$.

\subsubsection{Merging of distant isolated nodes}
An additional mechanism, which can be at work
for arbitrarily large $k_a$, involves the formation
of clusters resulting from the cumulative merging of massive distant
nodes, with no role played by topologically connected clusters.
The average distance between a node of degree $k$ and its
closest node with degree at least $k$ is given by~\cite{castellano2020cumulative}
\begin{equation}
  d(k) \simeq 1+a(\gamma)\ln \left (\frac{k}{\km} \right),
\end{equation}
where
\begin{equation}
a(\gamma)=\frac{\gamma-3}{\ln(\kappa)}
\end{equation}
and $\kappa = \left< k^2 \right>/ \left< k\right>-1$ is the network
branching factor.
If $r(k) \geq d(k)$, these two nodes merge in a single CMP cluster.
For this reason, if $r(k)$ grows with $k$ faster than $d(k)$,
all nodes with a degree larger than $k_x$, where $k_x$ is the solution of
the equation
\be
r(k_x)=d(k_x),
\label{kx}
\ee
will be part of the same CMP cluster and
\begin{equation}
S^{(2)}_{\text{CMP}} = \int_{k_x}^{\infty}dk P(k) = \left( \frac{k_x(k_a)}{\km} \right)^{1-\gamma}.
\label{eq:size_2}
\end{equation}
Notice that, since $r(m)$ depends on $k_a$, also $k_x$ is a function
of $k_a$.

If instead $r(k)$ grows, for large $k$, more slowly than $d(k)$ then
isolated massive nodes are too far away from each other and
this mechanism does not activate.
In such a case, since no mechanism is active for diverging $k_a$,
the CMP threshold is necessarily finite.

\subsubsection{Crossover between the two mechanisms}
If Eq.~\eqref{kx} has a solution then the CMP asymptotic regime for
large $k_a$ is described by Eq.~\eqref{eq:size_2}.  As we will see in
specific examples below, the asymptotic behavior may be preceded by an
interval of $k_a$ values where the extended DOP mechanism dominates. In
such a case the asymptotic regime is reached after a crossover at a
degree value $k_2^*$ given by the solution of the equation
\begin{equation}
S_{\text{CMP}}^{(1)}(k_2^*)=S_{\text{CMP}}^{(2)}(k_2^*).
\label{eq:crossover_k_2}
\end{equation}
Notice that, since Eq.~\eqref{eq:SCMP1} is an overestimate of the first
contribution due to the extended DOP mechanism, the solution of
Eq.~\eqref{eq:crossover_k_2} is actually an upper bound of the true
crossover scale $k_2^*$.

The picture presented above is valid in networks of infinite size. In
numerical simulations on finite networks the asymptotic regime can be
actually observed for large $k_a$ only if the system is large enough
that the maximum degree $k_{\text{max}}(N)$ is much larger than $k_2^*$.
For random graphs with $\gamma>3$, generated using the Uncorrelated
Configuration Model (UCM)~\cite{catanzaro2005generation}, the maximum
degree scales with $N$ as $k_{\text{max}} \sim N^{1/(\gamma-1)}$.
Hence, to observe the asymptotic regime it is necessary that
$N \gg N_2^*$, where $N_2^* = k_2^{*(\gamma-1)}$. For the particular
form of $r(m)$ describing SIS epidemic dynamics this value is much
larger than the sizes that can be simulated; as a consequence in
Ref.~\cite{castellano2020cumulative} only the preasymptotic extended DOP
regime was observed.  We will see below that for various forms of the
function $r(m)$ the truly asymptotic regime can be cleanly observed in
simulations.

The finite size of networks considered in simulations induces also the
presence of a size-dependent effective threshold $\kac(N)$ even if there
is no threshold in the limit of infinite network size. This must be kept in mind
when interpreting simulation results. See Appendix~\ref{FSS} for
details.

\section{Results for two specific forms of the interaction range}
\label{sec:IV}

In this Section we analyze in detail what happens for two specific
choices of the functional dependence of the interaction range $r(m)$ on
the mass.  In each case, after deriving the predictions of the scaling
theory we compare them with the results of numerical simulations of the
CMP process.  These were performed by considering random networks built
according to the UCM algorithm~\cite{catanzaro2005generation} with
minimum degree $\km=3$ and various sizes $N$. To avoid the strong
sample-to-sample fluctuations in the value of the maximum
degree~\cite{Boguna2009} we extracted the degree distribution imposing
the degrees to be strictly constrained between $\km$ and
$N^{1/(\gamma-1)}$.

\subsection{Algebraically growing interaction range}
\label{sec:algebr-grow-inter}


In this subsection we consider an interaction range growing algebraically
with the cluster mass
\begin{equation}
  r(m)=\left( \frac{m}{k_a}\right)^{\alpha},
\end{equation}
where $\alpha>0$ is a fixed parameter.  The case $\alpha=1$ corresponds
to the linear case studied in Ref.~\cite{castellano2020cumulative}.

The effectiveness of the extended DOP mechanism strongly depends on the
value of $\alpha$.  For $\alpha \geq 1$ a cluster of size 2 has an
interaction range at least equal to 2 and it can merge with another
cluster if at distance 2 from it.  Moreover, also isolated nodes with
$k>k_e=2^{1/\alpha} k_a$ and distance equal to 2 merge with DOP
clusters.  Instead if $\alpha<1$ the above statements are no longer
true, and only sufficiently massive clusters or isolated nodes at
distance $2$ may participate to merging events.  In such a case
Eq.~\eqref{eq:size_1} is an overestimation of the size of the CMP
largest cluster.

Concerning the second mechanism, since $r(m)$ grows algebraically, and
the average distance grows logarithmically, the interaction range is,
asymptotically, always larger than the distance.  Hence the equation
$r(k_x)=d(k_x)$ always has a solution, and the second mechanism is
active for sufficiently large degrees, no matter the value of $\alpha$.  Setting
$k_x=\omega k_a$, from Eq.~\eqref{kx} we have the transcendental
equation
\begin{equation}
  \omega^{\alpha} = 1+a(\gamma)\ln \left (\omega\right)+ \ln\left(\frac{k_a}{\km} \right)
  \label{eq:omega_power}
\end{equation}
which can be solved for $\omega$ as (see
Appendix~\ref{appendix:lambert})
\begin{equation}
  \omega(k_a)
  =\frac{e^{-\frac{1}{a(\gamma)}}}{k_a/\km}\exp{\left[-\frac{1}{\alpha}
      W_j \left(-\frac{\alpha
          e^{-\frac{\alpha}{a(\gamma)}}}{a(\gamma)}\left(\frac{k_a}{\km}\right)^{-\alpha}
      \right) \right]},
\label{eq:omega_exact_solution_W}
\end{equation}
where $W_j(z)$ is the Lambert $W$ or product logarithm
function~\cite{corless1996lambertw}. The branch corresponding to the
physical solution is the one with $j=-1$, since the branch with $j=0$
implies $\omega \to 0$ as $k_a \to \infty$, in contradiction with the
requirement that $k_x \ge k_a$.  Expanding $W_{-1}(z)$ for small
argument (that is for $k_a \gg \km$), we get, see
Appendix~\ref{appendix:lambert},
\begin{equation}
  \omega \simeq \left[1+\frac{a(\gamma)}{\alpha}\ln\left(\frac{
        a(\gamma)}{\alpha}\right)+a(\gamma)\ln \left(\frac{k_a}{\km}
    \right) \right]^{1/\alpha}.
\end{equation}
Thus we end up, neglecting constants and terms of lower order, with a CMPGC of size
\begin{equation}
  S_{\text{CMP}}^{(2)} \sim \left[a(\gamma)\ln \left(\frac{k_a}{\km}
    \right) \right]^{(1-\gamma)/\alpha}\left( \frac{k_a}{\km}
  \right)^{1-\gamma}.
\label{eq:size_2_power}
\end{equation}
Eq.~\eqref{eq:size_2_power} is in
agreement with the results in \cite{castellano2020cumulative}, which are
recovered setting $\alpha=1$. We see that the
introduction of an exponent $\alpha$ tuning the interaction range
does not change the critical exponent $\beta=1$, but only modifies
logarithmic corrections.

To calculate the crossover degree $k_2^*$, inserting
Eq.~\eqref{eq:SCMP1} and Eq.~\eqref{eq:size_2}, evaluated for $k_a=k_2^*$
into Eq.~\eqref{eq:crossover_k_2} we obtain, after some transformations,
\begin{equation}
  k_2^* = \km \left[\frac{\bar{\omega}^{1-\gamma}}{\left< k\right>}
  \right]^{\frac{1}{3-\gamma}},
\label{k23}
\end{equation}
where $\bar{\omega}$ is the solution of
\begin{equation}
  \bar{\omega}=\left[1+\frac{1}{\ln(\kappa)}\ln
    \left<k\right>+b(\gamma)\ln \left( \bar{\omega} \right)
  \right]^{1/\alpha},
  \label{omega}
\end{equation}
where we have defined
\begin{equation}
 b(\gamma)=\frac{2(\gamma-2)}{\gamma-3}a(\gamma)=\frac{2(\gamma-2)}{\ln(\kappa)}.
\end{equation}
The equation for $\bar{\omega}$ can be solved analytically by using the Lambert $W$
function with the same strategy developed in
Appendix~\ref{appendix:lambert}.
Alternatively, we can solve numerically Eq.~\eqref{omega} as a fixed point equation
for $\bar{\omega}$, and then insert this value in Eq.~\eqref{k23}.
\begin{figure}[t]
  \includegraphics[width=0.5\textwidth]{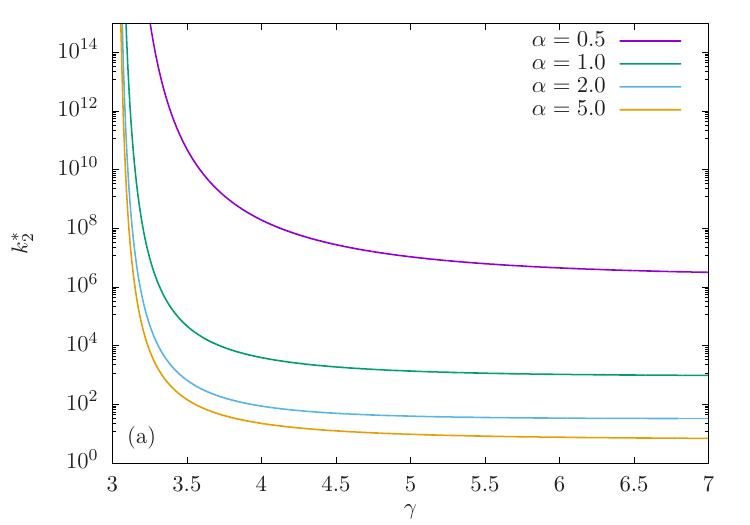}
  \includegraphics[width=0.5\textwidth]{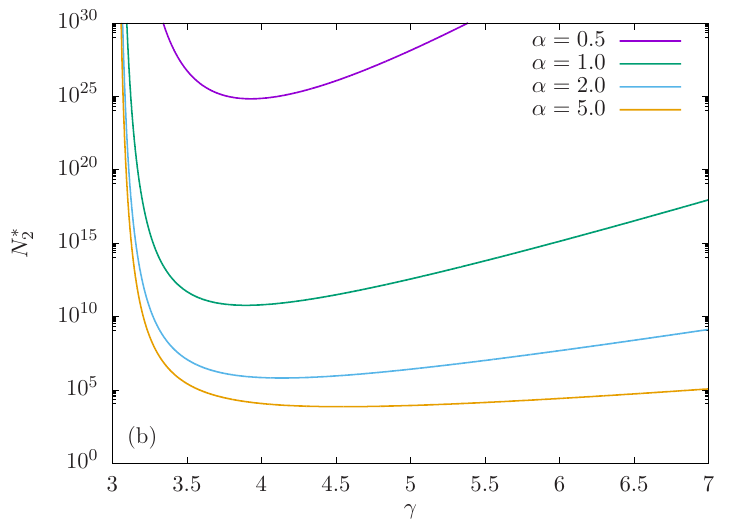}
  \caption{
    Analytical results for algebraically growing interaction range.
    Plot of $k_2^*$ (a) and $N_2^*$ (b) as a function of $\gamma$ for
    algebraically growing interaction range and several values of $\alpha$.}
  \label{fig:results_power}
\end{figure}
Fig.~\ref{fig:results_power} shows that the minimum graph size $N_2^*$
needed to observe the second mechanism at work is indeed much beyond values
that can be considered in practice when $\alpha \le 1$,
while it attains feasible values for $\alpha>1$ and $\gamma$ close to 4.

\begin{figure}[t]
  \centering
  \includegraphics[width=0.48\textwidth]{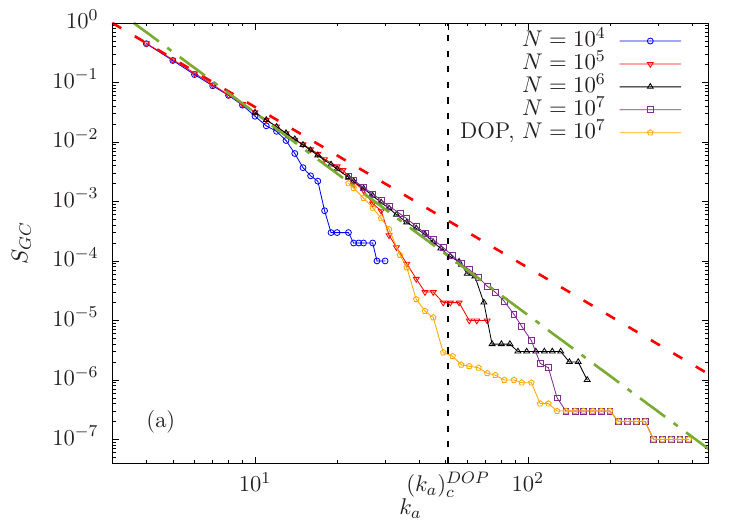}
  \includegraphics[width=0.48\textwidth]{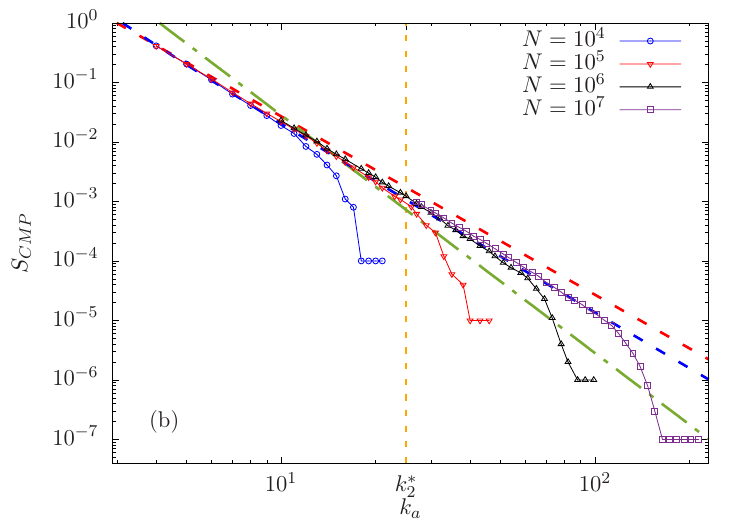}
  \caption{Comparison of analytical and simulation results for the size
    of the CMP largest cluster as a function of $k_a$, for algebraically
    growing interaction range and different combinations of $\gamma$ and
    $\alpha$ values: (a) $\gamma=3.7$ and $\alpha=0.5$; (b) $\gamma=4$ and
    $\alpha=5$. The red dashed line is the scaling with
    exponent $1-\gamma$, the green dot-dashed line is the scaling with
    exponent $2(2-\gamma)$ and the blue dashed line is the prediction of
    $S_{\text{CMP}}^{(2)}$ given by Eq.~\eqref{eq:size_2} where $k_x=\omega k_a$
    and $\omega$ is given by Eq.~\eqref{eq:omega_exact_solution_W}.
    In panel (a) we also report the results of a simulation of the DOP
    process on the network with size $N=10^7$.}
  \label{alpha}
\end{figure}

Numerical simulations of the CMP process (see Fig.~\ref{alpha}) confirm
the analytical predictions.  For $\alpha=0.5$ and $\gamma=3.7$ the
crossover to the asymptotic behavior happens at $N_2^*\sim 10^{25}$
($k_2^*$ is of the order of $10^{10}$).  We thus have only access to the
first regime, dominated by the extended DOP mechanism. Indeed, as
predicted by Eq.~\eqref{eq:SCMP1}, the order parameter $S$ decays with
an exponent $2(2-\gamma)$ that extends well beyond the DOP threshold.
For $\alpha=5$ and $\gamma=4$ instead the crossover value is predicted
to be $k_2^*\approx 23$.
Correspondingly we observe, for $N \gg N_2^* \approx 10^4$, the asymptotic decay to be well
described by Eq.~\eqref{eq:size_2_power}. Notice that in this case also
the logarithmic correction is necessary to match the decay.

\subsection{\label{sec:V}Logarithmically growing interaction range}
Another interesting class of CMP processes is the one described by an
interaction range growing logarithmically with the mass, that is
\begin{equation}
  r(m)=1 + \delta\ln\left( \frac{m}{k_a}\right),
\end{equation}
where $\delta>0$ is a parameter tuning how fast the range increases.

In this case, since the interaction range grows only logarithmically
with the cluster mass, the asymptotic ineffectiveness of the extended
DOP mechanism is expected to be even more severe than in the previous
case for $\alpha<1$.  Concerning the merging of massive isolated distant
nodes, since in this case both $r(m)$ and $d(k)$ grow logarithmically,
it is not always true that a degree $k_x$ exists such that
$r(k) \geq d(k)$ for $k \ge k_x$.  This condition holds when
\begin{equation}
  \delta \ln\left(\frac{k}{k_a} \right) \geq a(\gamma)\ln\left(\frac{k}{k_a} \right)
  + a(\gamma)\ln\left(\frac{k_a}{\km} \right).
\end{equation}
This implies that $k_x$ exists only if
\begin{equation}
 \delta > a(\gamma).
\end{equation}

\begin{figure}[t]
  \centering
  \includegraphics[width=0.47\textwidth]{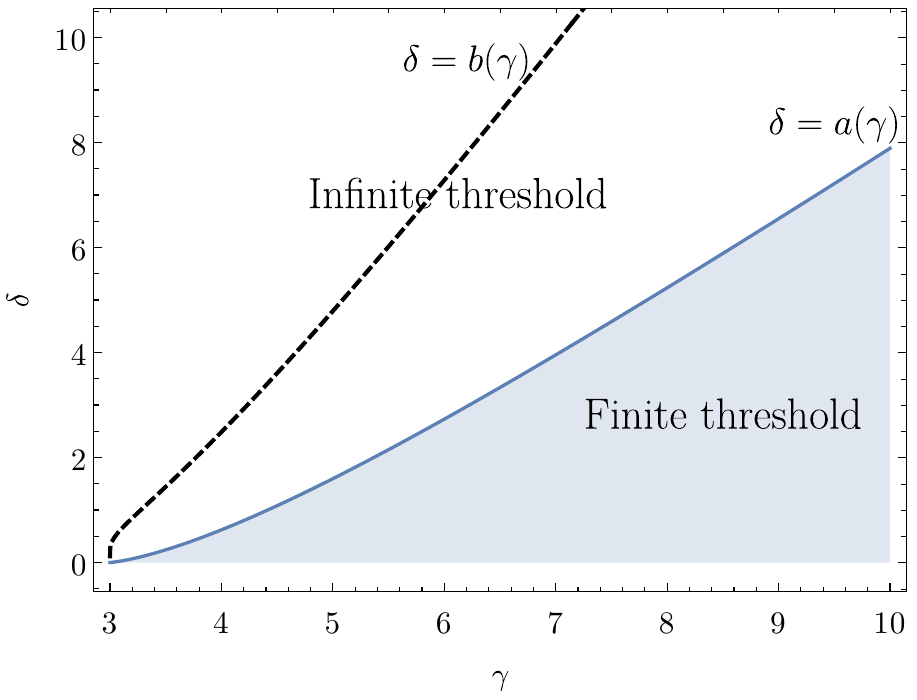}
  \caption{Phase diagram of CMP with logarithmically growing
      interaction range in the $(\delta,\gamma)$ plane.  In the
    shadowed region below the blue solid line, the second
    mechanism does not activate at all. In the region between the
      solid and dashed lines the second mechanism activates, but it
      is subleading compared to the first mechanism (as long as the
      first mechanism is at work). Above the dashed
    line the second mechanism activates and is leading with respect to
    the extended DOP mechanism.}
  \label{phase_diagram_log}
\end{figure}

This result indicates a completely different phenomenology from the one
found in Section~\ref{sec:algebr-grow-inter}. The critical line
$\delta=a(\gamma)$ divides the $(\delta,\gamma)$ plane in two regions
(see Fig.~\ref{phase_diagram_log}): For $\delta>a(\gamma)$, the
mechanism responsible for the merging of distant isolated nodes of large
degree is active for large $k_a$; below $\delta=a(\gamma)$ instead, the
interaction range grows too slowly with respect to the average distance
between nodes of degree $k$ and hence the merging of all massive
isolated nodes in a single CMP cluster does not occur.  As a
consequence, we can argue that for $\delta>a(\gamma)$ the CMP has an
infinite threshold, while for $\delta<a(\gamma)$ the order parameter
$S_{\text{CMP}}$ must go to zero at some finite critical value $\kccmp$.

In the region $\delta>a(\gamma)$, we can solve the equation for
$\omega=k_x/k_a$ and compute the size $S_{\text{CMP}}^{(2)}$. We have
\begin{equation}
  \delta \ln(\omega )=a(\gamma) \ln (\omega)+ a(\gamma)\ln\left(\frac{k_a}{\km}\right),
\end{equation}
from which it follows
\begin{equation}
k_x = \omega k_a = \left(\frac{k_a}{\km} \right)^{a(\gamma)/\left[\delta-a(\gamma)\right]} k_a.
\end{equation}
Inserting the last expression into Eq.~\eqref{eq:size_2}
\begin{equation}
S^{(2)}_{\text{CMP}} \simeq \left(\frac{k_a}{k_\text{min}}\right)^{\frac{(1-\gamma)\delta}{\left[\delta-a(\gamma)\right]}}.
  \label{S2l}
\end{equation}
Hence we find that the critical exponent
\be
\beta=\frac{\delta}{\delta-a(\gamma)}
\ee
(see Eq.~\eqref{eq:3}) is a continuously changing function of the parameters $\gamma$ and $\delta$.

\begin{figure}[t]
  \includegraphics[width=0.44\textwidth]{./exp}
  \includegraphics[width=0.4\textwidth]{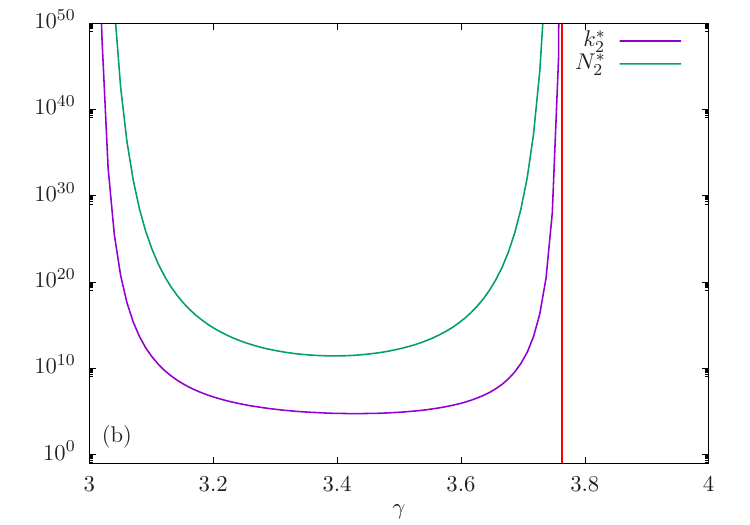}\\
  \includegraphics[width=0.4\textwidth]{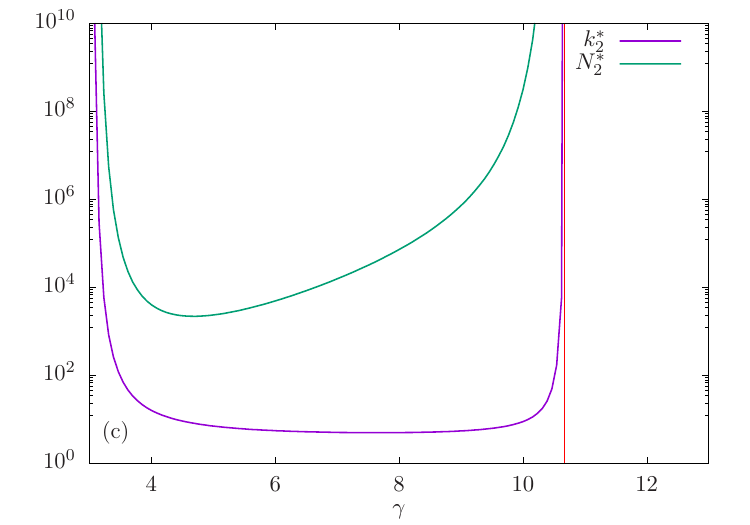}
  \caption{Analytical results for logarithmically growing interaction range.
  (a) Three-dimensional representation of
  $\mu(\gamma, \delta)$ of Eq.~\eqref{exponent_mu}.
    (b) Plot of $k_2^*$ and $N_2^*$ as a function of $\gamma$ for $\delta=2$.
    (c) Plot of $k_2^*$ and $N_2^*$ as a function of $\gamma$ for $\delta=20$.
    In both panels b) and c) the condition $\delta>b(\gamma)$ is verified.
    The red vertical lines represent the value of $\gamma$ for which
    $\delta=b(\gamma)$ and thus $k_2^*$ and $N_2^*$
    diverge.}
  \label{fig:k2}
\end{figure}

Also Eq.~\eqref{eq:crossover_k_2} for the crossover scale $k_2^*$ can
be exactly solved in this case.  Inserting the expressions~\eqref{eq:SCMP1}
and~\eqref{S2l} into Eq.~\eqref{eq:crossover_k_2} and performing
straighforward calculations we obtain
\begin{equation}
  k_2^*=\km \langle k \rangle^{\mu(\gamma, \delta)},
  \label{k22}
\end{equation}
where
\begin{equation}
  \mu(\gamma, \delta)=\frac{\delta-a(\gamma)}{(\gamma-3)\left[\delta-b(\gamma)\right]}.
  \label{exponent_mu}
\end{equation}
See Fig.~\ref{fig:k2} for a plot of the function $\mu(\gamma,\delta)$.

The expression for $k_2^*$ in Eq.~\eqref{k22} actually applies only for
$\delta>b(\gamma)$.  Indeed, as discussed above, for $\delta<a(\gamma)$
the merging of distant nodes is not active.  Hence the CMP model is
practically identical to DOP and we expect $\kccmp \approx \kcdop$.  For
$a(\gamma)<\delta<b(\gamma)$, instead, the second mechanism activates
asymptotically but it is subleading with respect to the first.  In this
case the extended DOP mechanism dominates for small values of $k_a$ but,
being asymptotically ineffective, at some point the second mechanism
takes over.  This crossover does not occur at the $k_2^*$ predicted by
Eq.~\eqref{k22} (which is smaller than $\km$ in this case) but where the
extended DOP mechanism stops working.  For $\delta>b(\gamma)$, we have
at $k_2^*$ a true crossover between the two mechanisms: The merging of
distant isolated nodes governs the asymptotic behavior of
$S_{\text{CMP}}$.  As we can see in Fig.~\ref{fig:k2}, plotting $k_2^*$
and $N_2^*$ as a function of $\gamma$ in the region $\delta>b(\gamma)$
shows that for small values of $\delta$ the crossover is largely out of
reach in simulations; instead for larger values of $\delta$, $N_2^*$ is
strongly reduced and hence it is possible to observe the asymptotic
regime in simulations.

\begin{figure}[h!]
  \centering
  \includegraphics[width=0.4\textwidth]{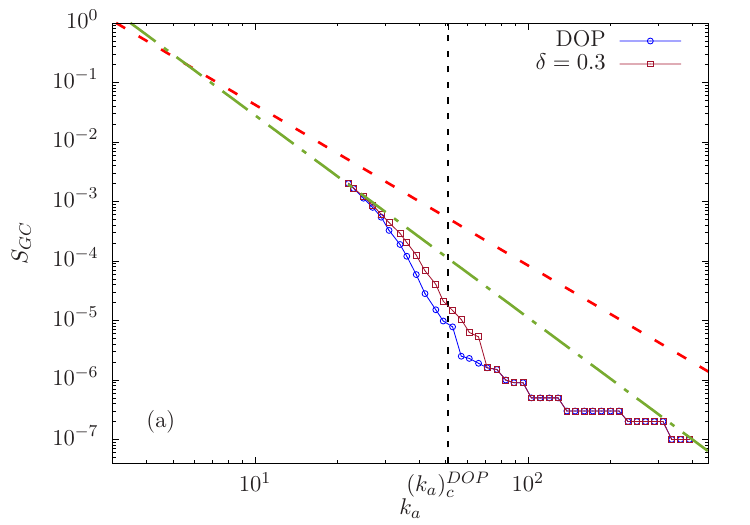}\\
  \includegraphics[width=0.4\textwidth]{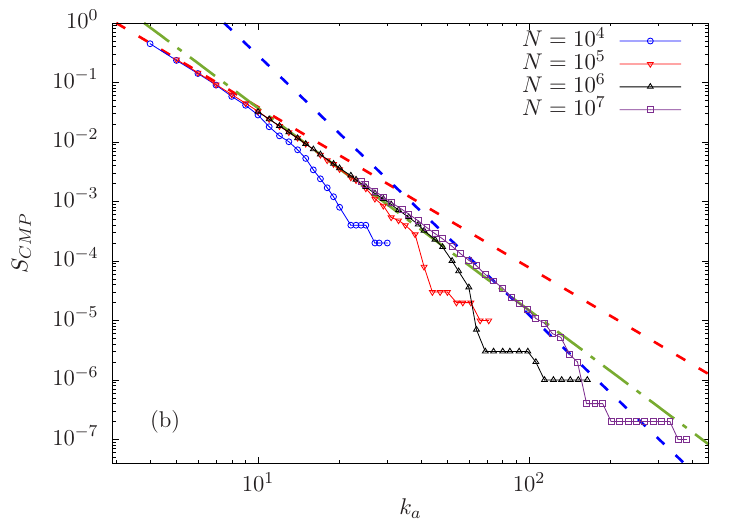}\\
  \includegraphics[width=0.4\textwidth]{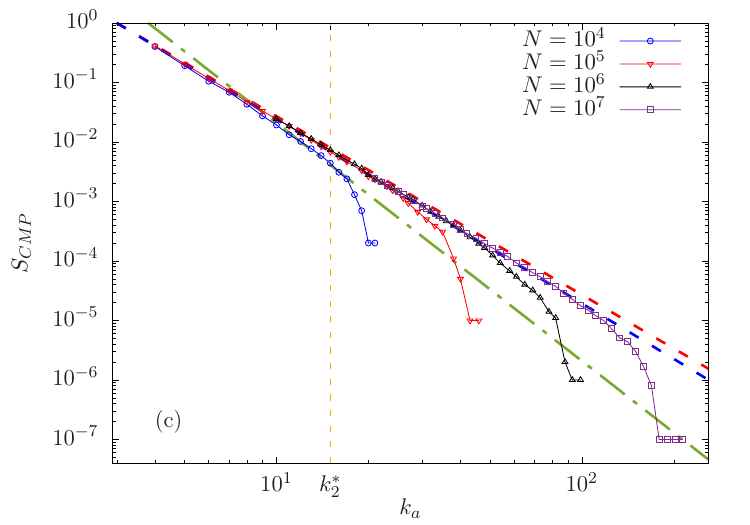}
  \caption{Comparison of analytical and simulation results for
      the size of the CMP largest cluster as a function
      of $k_a$, for logarithmically growing interaction range and
      different combinations of $\gamma$ and $\alpha$ values:
  (a) $\gamma=3.7$ and $\delta=0.3$,
    so that $\delta<a(\gamma)$, the system size is $N=10^7$; (b) $\gamma=3.7$ and $\delta=1$, so
    that $a(\gamma)<\delta<b(\gamma)$; (c) $\gamma=4$ and $\delta=20$,
    so that $\delta>b(\gamma)$.  The vertical dashed line in panel (c)
    is the value of $k_2^*$. The red dashed line is the scaling with exponent
    $1-\gamma$, the green dot-dashed line is the scaling with exponent
    $2(2-\gamma)$ and the blue dashed line is the prediction of
    $S_{\text{CMP}}^{(2)}$ given in Eq.~\eqref{S2l}.}
  \label{fig:results_log}
\end{figure}

Results of numerical simulations, reported in
Fig.~\ref{fig:results_log}, confirm this overall picture.  In
Fig.~\ref{fig:results_log}(a), corresponding to $\delta<a(\gamma)$, the
behavior of CMP practically coincides with that of DOP.  For
$a(\gamma)<\delta<b(\gamma)$, Fig.~\ref{fig:results_log}(b), the
extended DOP mechanism dominates, up to a finite threshold. The true
asymptotic behavior here is the one predicted by Eq.~\eqref{S2l}, but it
would be observed only for much larger system size $N$.  For
$\delta>b(\gamma)$ instead, $S_{\text{CMP}}$ nicely follows the
asymptotic prediction of Eq.~\eqref{S2l}, after the crossover scale.

Finally, let us point out that the behavior of the CMP model as
$\delta \to a(\gamma)^-$ is nontrivial.  Let us consider a fixed value
of $\gamma$. For $\delta=0$ the DOP process has a finite critical point
$\kcdop$. For $\delta>a(\gamma)$, we have instead an infinite critical
point and a CMPGC always exists. What happens in the intermediate
region?  How does the finite critical point $\kccmp$ change in the
region as a function of $\delta$?  We know that $\kccmp \to \kcdop$ as
$\delta \to 0$, but what happens for $\delta \to a(\gamma)^{-}$?  As
long as $\delta<a(\gamma)$ the merging of distant isolated nodes is not
at work for large degrees. Hence the transition is governed by the
extended DOP mechanism and occurs not far from the DOP critical point.
As a consequence, we expect $\kccmp$ to be a discontinuous function of
$\delta$ at fixed $\gamma$, jumping from a finite value to $\infty$ when
$\delta$ reaches $a(\gamma)$. A direct numerical verification of this
conjecture is however impossible, due to finite-size effects.

\section{\label{sec:VI}Discussion and conclusions}

In this paper, extending the work of
Refs.~\cite{menard2016percolation,castellano2020cumulative}, we
introduced a general formulation for a new class of percolation
processes in networks, dubbed cumulative merging percolation (CMP),
characterized by the fact that nodes belonging to a cluster need not be
topologically connected. Clusters are instead defined via an iterative
procedure that may merge network subsets even if they are far apart in
the network. In this sense CMP is a long-range percolation process,
qualitatively different from extended-neighborhood processes, which have
only a finite interaction range.  We then considered a specific subclass
of CMP models, characterized by node masses equal to node degrees
and a degree-ordered activation of nodes.
This class generalizes the CMP model introduced in
Ref.~\cite{castellano2020cumulative} by allowing for an arbitrary
(non-decreasing) functional dependence $r(m)$ of the interaction range
on the cluster mass. Building on~\cite{castellano2020cumulative} we
developed a scaling theory for this class of CMP models on power-law
degree-distributed networks, which allows us to determine the behavior of
the order parameter $S_{\text{CMP}}$ and the associated critical
properties.

We then considered two specific functional forms for $r(m)$.  We first
focused on the case in which $r(m)$ grows algebraically with $m$.  We
showed that a giant cluster (CMPGC) always exists for any value of $k_a$ (the
control parameter that sets the degree threshold for node activation),
even if the interaction range grows sublinearly, and the critical
exponent $\beta$ is the same of the linear case. Furthermore, for proper
choices of model parameters, we were able to actually observe the
crossover to the true asymptotic regime in numerical simulations, in
perfect agreement with the theoretical predictions.  Note that this
observation is not possible with the linear interaction range used in
Ref.~\cite{castellano2020cumulative}, since it would require network
sizes out of reach for numerical simulations.  We then considered a
logarithmically growing interaction range, in order to study the
nontrivial competition between the distance among active nodes and their
interaction range itself. We discovered that a CMPGC exists for
arbitrarily large $k_a$ only if the interaction range grows ``fast
enough'' with respect to the distance. We identified a critical line in
the parameters space that separates a region in which a CMPGC always
exists from a region in which a CMPGC exists only up to a finite
critical point $\kccmp$.

Many aspects of the present work are worth future exploration.
Indeed the CMP process allows for countless variations that may give rise to
new nontrivial critical phenomena.  For instance, what happens if CMP is
realised when nodes are activated at random, rather than in a
degree-ordered way?  What changes with a different initial assignment of
node masses, i.e., $m_i \ne k_i$?
What if the mass of a cluster is given by the product, rather than the
sum of individual masses?
The investigation of these 
and other models described by other choices of the CMP parameters is an
interesting task for future research.
\change{Another possible avenue for future investigations is the exploration
of connections between generic forms of Cumulative Merging Percolation
and epidemic processes. For example, CMP with algebraically growing
interaction range can be seen as a description of a suitably defined
SIS model on uncorrelated weighted networks.}

\appendix

\section{Finite-size effects}
\label{FSS}
In the cases where the threshold is infinite (i.e., there is a
giant cluster for any $k_a$),
the finiteness of the network induces the existence of a finite
size-dependent threshold $\kac(N)$, that diverges as $N$ grows.
Its detailed behavior depends on which mechanism dominates when $k_a$
reaches the value $k_{\text{max}}(N)$.

If $k_{\text{max}}(N)<k_2^*$, finite-size effects appear during
the preasymptotic regime where the extended DOP mechanism rules. The
effective threshold is thus given by the condition
$k_c=k_{\text{max}}(N)$, implying
\begin{equation}
  \kac=\km k_{\text{max}}^{1/(\gamma-2)}=\km N^{1/[(\gamma-1)(\gamma-2)]}.
  \label{finite_size_pow_2}
\end{equation}

If instead $k_{\text{max}}(N)>k_2^*$, finite-size effects appear when the
CMPGC is governed by the second mechanism.
Thus the asymptotic
behavior ends when $k_x=k_{\text{max}}(N)$, implying
\begin{equation}
  k_x[\kac]=N^{1/(\gamma-1)}.
  \label{finite_size_passaggio}
\end{equation}

\begin{figure}[t]
  \centering
  \includegraphics[width=0.48\textwidth]{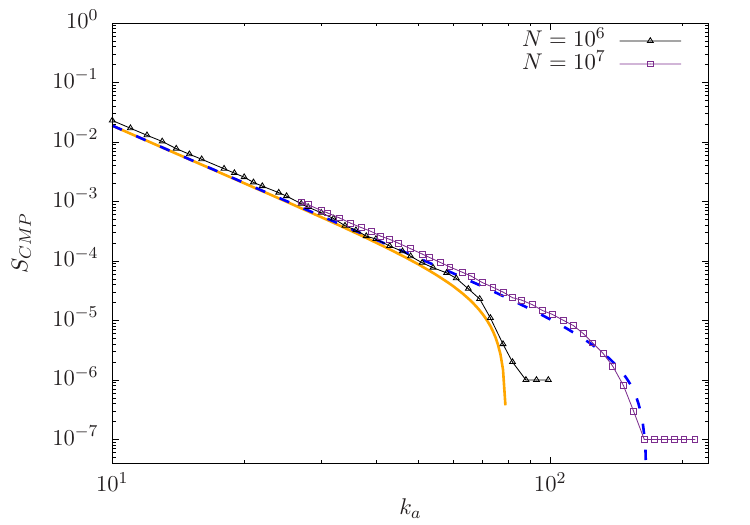}
  \caption{\change{Comparison of analytical and simulation results for the size
    of the CMP largest cluster as a function of $k_a$, for algebraically
    growing interaction range, $\gamma=4$ and $\alpha=5$.
    Symbols are the results of numerical simulations in networks of size
    $N=10^6$ (black triangles) and $N=10^7$ (purple squares).
    Lines are the predictions given by Eq.~\eqref{eq:size_2}
    where $k_x=\omega k_a$,
    $\omega$ is given by Eq.~\eqref{eq:omega_exact_solution_W} and
    the integral is performed only up to $k_{\text{max}}(N)$.
    The solid orange line is for $N=10^6$, the blue dashed line is for
    $N=10^7$.}}
  \label{alphaFS}
\end{figure}

\change{A precise prediction of how the order parameter
$S_{\text{CMP}}^{(2)}$ is cut off when $k_x$ approaches $k_{\text{max}}(N)$
is obtained by performing the integral in Eq.~\eqref{eq:size_2}
only up to $k_{\text{max}}(N)$ (see Fig.~\ref{alphaFS}).}

For $k_a>\kac$ neither of the two mechanisms is at work and
$S_{\text{CMP}}$ rapidly goes to zero~\footnote{Of course, it does not
go to zero in a finite $N$ simulation because it tends to $1/N$, which
is the minimum value for $S_{\text{CMP}}$.}.

\section{Solution of transcendental equations Eq.~\eqref{eq:omega_power}}
\label{appendix:lambert}

Let us consider the general equation
\begin{align}
  x = A + B e^{Cx},
  \label{equation_appendix}
\end{align}
where $A$, $B$ and $C$ are complex numbers. Subtracting $A$ on both sides and
multiplying by $C$ we get
\begin{equation}
  C(x-A)=BCe^{Cx}.
\end{equation}
Setting $t=C(x-A)$ and multiplying by $-e^{-t}$ we
get
\begin{equation}
  -te^{-t}= -BCe^{AC}.
  \label{eq:1}
\end{equation}
Eq.~\eqref{eq:1} can be solved using the Lambert $W$ or product
logarithm
function, defined as the function fulfilling the
expression~\cite{corless1996lambertw}
\begin{equation}
  \label{eq:4}
  W(z) e^{W(z)} = z.
\end{equation}
The Lambert $W$ function can be considered as the inverse of the
function $f(z) = z e^{z}$, in such a way that
\begin{equation}
  \label{eq:5}
  W(z e^z) = z.
\end{equation}
The function $f(z) = z e^{z}$ is not invertible for every $z$, and
therefore $W(z)$ is multivalued and has several branches, $W_j(z)$. For
real $w$ and $z$, the equation $w = z e^z$ can be shown to have only two
branches, $W_0(z)$, the so-called \textit{principal branch}, and
$W_{-1}(z)$. In this case, it is easy to prove that
\begin{itemize}
\item for $z>0$ there exist only one solution $w=W_0(z)$;
\item for $-e^{-1}\leq z \leq 0$ there are two solutions corresponding
  to the two branches $W_0(z)$ and $W_{-1}(z)$;
\item for $z<-e^{-1}$ there is not any solution.
\end{itemize}

Now, applying $W(z)$ to both sides of Eq.~\eqref{eq:1}, we obtain
\begin{equation}
  \label{eq:6}
  W\left(  -te^{-t} \right) = -t = W \left( -BCe^{AC} \right),
\end{equation}
which, resolving for $x = A + t/C$, leads to the solution for
Eq.~\eqref{equation_appendix}
\begin{align}
  x=A-\frac{1}{C}W_j\left(-BCe^{AC} \right).
  \label{lambert_solution}
\end{align}
If we are interested in real solutions, must require that
$z \geq -e^{-1}$, that is
\begin{align}
  BCe^{AC+1}\leq 1.
\label{lambert_condition}
\end{align}
In order to approximate the Lambert $W$ function, we can use the
expansions for the real branches~\cite{corless1996lambertw}
\begin{align}
  W_0(z)=\sum_{n=1}^{\infty}\frac{(-n)^{n-1}}{n!}z^n=z-z^2+\frac{3}{2}z^3-...
\end{align}
and
\begin{align}
  W_{-1}(z)&=L_{1}-L_{2}+\sum _{l=0}^{\infty}\sum _{m=1}^{\infty}{\frac {(-1)^{l}\left[{\begin{smallmatrix}l+m\\l+1\end{smallmatrix}}\right]}{m!}}L_{1}^{-l-m}L_{2}^{m} \nonumber\\
           &=L_{1}-L_{2}+\frac{L_{2}}{L_{1}}+...
\end{align}
where $L_{1}=\ln(-z)$ and $L_2=\ln\left[-\ln(-z) \right]$, and
$z \to 0^{-}$

Turning finally to Eq.~\eqref{eq:omega_power}, namely
\begin{equation}
  \omega^{\alpha} = 1+a(\gamma)\ln \left (\omega\right)+
  \ln\left(\frac{k_a}{\km} \right),
  \label{eq:omega_power_app}
\end{equation}
if we define $\omega = e^x$, i.e. $x = \ln(\omega)$, we can write
Eq.~\eqref{eq:omega_power_app} in the form
\begin{equation}
  \label{eq:7}
  x = -\frac{1}{a}\left( 1 + \ln\left( \frac{k_m}{\km} \right) \right) +
  \frac{1}{a}e^{\alpha x}.
\end{equation}
This equation takes the exact form of Eq.~\eqref{equation_appendix} if
we define
\begin{equation}
  \label{eq:8}
  A = -\frac{1}{a}\left( 1 + \ln\left( \frac{k_m}{\km} \right) \right),
  \; \; B = \frac{1}{a}, \;  \; C = \alpha.
\end{equation}
The solution of Eq.~\eqref{eq:7} is thus immediately given by
Eq.~\eqref{lambert_solution}, and from here, reverting the change
$\omega = e^x$, we recover the solution
Eq.~\eqref{eq:omega_exact_solution_W}.

\begin{acknowledgments}
  R. P.-S. and C. C. acknowledge financial support from the Spanish
  MCIN/AEI/10.13039/501100011033, under Project
  No. PID2019-106290GB-C21.
  \change{G. C. and C. C. thank the project ``Complexity in Epidemiology'' of
  the Centro Ricerche Enrico Fermi.}
\end{acknowledgments}

%

\end{document}